\documentclass[aps,prl,twocolumn,superscriptaddress]{revtex4}

\usepackage[dvips]{graphicx}
\usepackage{ulem}
\usepackage{bm}
\renewcommand{\vec}[1]{\boldsymbol{#1}}

\begin{document}

\title{Evidence for gapped spin-wave excitations 
in the frustrated Gd$_2$Sn$_2$O$_7$ pyrochlore antiferromagnet from low-temperature specific heat measurements}

\author{J.~A.~Quilliam}
\author{K.~A.~Ross}
\affiliation{Department of Physics and Astronomy and Guelph-Waterloo Physics Institute, University of
Waterloo, Waterloo, ON N2L 3G1 Canada}
\affiliation{Institute for Quantum Computing, University of
Waterloo, Waterloo, ON N2L 3G1 Canada}
\author{A.~G.~Del~Maestro}
\affiliation{Department of Physics, Harvard University, Cambridge,
Massachusetts, 02138, USA}

\author{M.~J.~P.~Gingras}
\affiliation{Department of Physics and Astronomy and Guelph-Waterloo Physics Institute, University of
Waterloo, Waterloo, ON N2L 3G1 Canada}
\affiliation{Canadian Institute for Advanced Research,
180 Dundas Street West, Suite 1400
Toronto, ON M5G 1Z8
Canada}
\affiliation{Department of Physics and Astronomy, University of
Canterbury, Private Bag 4800, Christchurch, New Zealand}

\author{L.~R.~Corruccini}
\affiliation{Physics Department, University of California-Davis,
Davis, CA 95616, USA}

\author{J.~B.~Kycia}
\affiliation{Department of Physics and Astronomy and Guelph-Waterloo Physics Institute, University of
Waterloo, Waterloo, ON N2L 3G1 Canada}
\affiliation{Institute for Quantum Computing, University of
Waterloo, Waterloo, ON N2L 3G1 Canada}

\date{\today}

\begin{abstract}

We have measured the low-temperature specific heat of the geometrically frustrated pyrochlore Heisenberg antiferromagnet Gd$_2$Sn$_2$O$_7$ in zero magnetic field. The specific heat is found to drop exponentially below approximately 350 mK.  This provides evidence for a gapped spin-wave spectrum due to an anisotropy resulting from single ion effects and long-range dipolar interactions.  The data are well fitted by linear spin-wave theory, ruling out unconventional low-energy magnetic excitations in this system, and allowing a determination of the pertinent exchange interactions in this material.

\end{abstract}


\maketitle


In the insulating R$_2$M$_2$O$_7$ magnetic pyrochlores, the rare-earth ions (R$^{3+}$) sit on a lattice of corner-sharing tetrahedra, resulting in geometrical frustration for antiferromagnetic nearest-neighbor exchange between Heisenberg spins.  As such, these materials display a wide variety of interesting and unusual behaviors.   Examples of observed phenomenology include spin ice~\cite{Ehlers2004}, spin glass~\cite{Gaulin1992,Dunsiger1996}, and spin liquid~\cite{Gardner1999} behaviors as well as long-range magnetic order~\cite{Bonville2003,Stewart2004,Bonville2004,Reotier2004-JPC,Yaouanc2005,Wills2006,Dunsiger2006} and perhaps some novel type of hidden order~\cite{Hodges2002}.  A common trend observed in all of these systems is an apparent persistence of spin dynamics down to the lowest temperatures as revealed by muon spin relaxation 
($\mu$SR)~\cite{Dunsiger1996,Bonville2004,Gardner1999,Dunsiger2006,Hodges2002,Yaouanc2005,Reotier2004-JPC}, M\"ossbauer spectroscopy~\cite{Bertin2002} and neutron spin echo~\cite{Ehlers2004} experiments.  The origin of persistent spin dynamics (PSDs) is a major open question in the study of highly frustrated insulating magnetic oxide materials~\cite{Bonville2003}.

Perhaps what is most perplexing, is that PSDs have been found in pyrochlores, such as Gd$_2$Ti$_2$O$_7$ (GTO)~\cite{Dunsiger2006,Yaouanc2005} and Gd$_2$Sn$_2$O$_7$ (GSO)~\cite{Bertin2002,Bonville2004,Reotier2004-JPC}, which, according to neutron scattering experiments~\cite{Stewart2004,Wills2006}, display long-range magnetic order below a critical temperature $T_c=0.74$~K for GTO~\cite{Stewart2004} and $T_c=1.0$~K for GSO~\cite{Wills2006}. This is highly unusual since conventional wisdom suggests that collective magnon-like excitations, hence spin dynamics, should freeze out in the limit of zero temperature.   Similarly, the Gd$_3$Ga$_5$O$_{12}$ antiferromagnetic garnet (GGG), which displays rather extended magnetic correlations below 140 mK~\cite{Petrenko1998}  that are now theoretically rationalized~\cite{Yavorskii2006}, also exhibits PSDs below 100 mK~\cite{Dunsiger2000}.  It is tempting to speculate that the unusual low-energy excitations giving rise to PSDs may result from a remnant of the high frustration that these systems possess and the lack of long-range order that would occur were it not for perturbatively small exchange interactions beyond nearest-neighbor~\cite{Reimers1991}, dipolar interactions~\cite{Palmer2000,Maestro2007} or single-ion anisotropy~\cite{Bramwell1994}.

We see Gd$_2$Sn$_2$O$_7$ (GSO) as a key system to investigate in seeking to convincingly demonstrate, via  measurements of the low temperature magnetic specific heat, $C_m(T)$, the existence of unconventional low-energy excitations, as suggested by the observation of PSDs.  
Previous $C_m$ measurements on this material between 350 mK and 800 mK (see Fig. 1) were found to be parametrized by a $C_m(T)\sim T^2$ law~\cite{Bonville2003}.  Such a temperature dependence of $C_m$ is unusual since conventional antiferromagnetic magnon excitations without a gap lead to $C_m(T)\sim T^3$ or, with an anisotropy energy gap $\Delta$, to $C_m(T)\sim\exp(-\Delta/T)$.
Such an unconventional $C_m(T)$ behavior further argues for the existence of unusual low-energy excitations in GSO. 
Interestingly, the related GTO, which also exhibits PSDs, displays $C_m(T)\sim T^2$ for 100 mK $\lesssim T \lesssim$ 500 mK~\cite{Yaouanc2005}. 

In contrast, a recent theoretical work~\cite{Maestro2007} argues that the lowest temperature $\sim 350$ mK considered in Ref.~[\onlinecite{Bonville2003}] for GSO corresponds to the {\it upper} temperature limit above which multi-magnon excitations start to proliferate, obscuring the genuine 
single-magnon/low-temperature regime.  In particular, calculations using the microscopic spin Hamiltonian~\cite{Palmer2000,Wills2006,Maestro2007} that describes the experimentally observed ground state of GSO~\cite{Wills2006} predict that $C_m$ should begin to drop away exponentially exactly at or just below $~350$ mK as gapped magnetic excitations become quenched at lower temperature~\cite{Maestro2007}. 

In this letter we present an investigation of the nature of the low-energy spin excitations in GSO through measurements of $C_m(T)$ down to 115 mK.  The results show an exponentially dropping specific heat at low temperature and, when compared with theoretical calculations~\cite{Maestro2007}, confirm the picture of conventional gapped antiferromagnetic magnons in Gd$_2$Sn$_2$O$_7$.


Polycrystalline samples of Gd$_2$Sn$_2$O$_7$ were prepared by solid state reaction as in Ref.~\cite{Luo2001}.  Gd$_2$O$_3$ and SnO$_2$ were mixed in stoichiometric ratio, pressed into pellets, and then fired in air for 48 hours at 1350$^\circ$C.  The pellets were then reground and the process repeated.  Powder X-ray diffraction spectra taken after the two firings were indistinguishable.  All peaks were indexed to space group $Fd\bar{3}m$, with the size $a$ of the cubic unit cell $a=10.460$ \AA, with no impurity peaks detectable at the 1\% level. High temperature susceptibility measurements yield a magnetic moment  of $7.95\mu_B$ for Gd$^{3+}$ in GSO, close to the expected free-ion moment of $7.94\mu_B$.

Specific heat measurements were performed using the quasi-adiabatic technique.  
 A 1~k$\Omega$ RuO$_2$ resistor was used as thermometer and calibrated to a commercially 
calibrated germanium resistance thermometer. A 10~k$\Omega$ metal-film resistor was used as a heater.  The sample was suspended from thin nylon threads and the heater and thermometer were fixed directly to the sample.  Leads to the thermometer and heater were made from 6 $\mu$m diameter, 1~cm long NbTi wires.  The sample was weakly heat-sunk to the mixing chamber of a $^3$He/$^4$He dilution refrigerator with Pt$_{0.92}$W$_{0.08}$ wire, chosen for its insignificant contribution to the addendum.
The addendum, due to other components such as heater and thermometer, was determined to be less than 2\% of the sample's heat capacity in the worst case at around 120 mK and thus does not effect the results of our analysis.

Slow thermal relaxation within the sample was observed (with a relaxation time constant $\tau_1 \sim 120$~s at 200 mK).  The thermal link was chosen so that the relaxation time of the sample temperature to cryostat temperature, $\tau_2$, would be much longer than $\tau_1$, thereby minimizing the temperature gradient within the sample.  A double exponential form was fit to the sample's temperature as a function of time after a heat pulse and the longer exponential was extrapolated to zero time, giving a measure of the sample's heat capacity.  The data presented here was taken using a long time constant $\tau_2\simeq 20\tau_1$.  Another experiment performed with $\tau_2\simeq 5\tau_1$ resulted in a slightly noisier measurement. Yet, the specific heat in both measurements overlapped within the estimated error bars of the data, ruling out significant inaccuracies due to the slow thermal relaxation in the sample.

\begin{figure}
\includegraphics[width=3.375in,keepaspectratio=true]{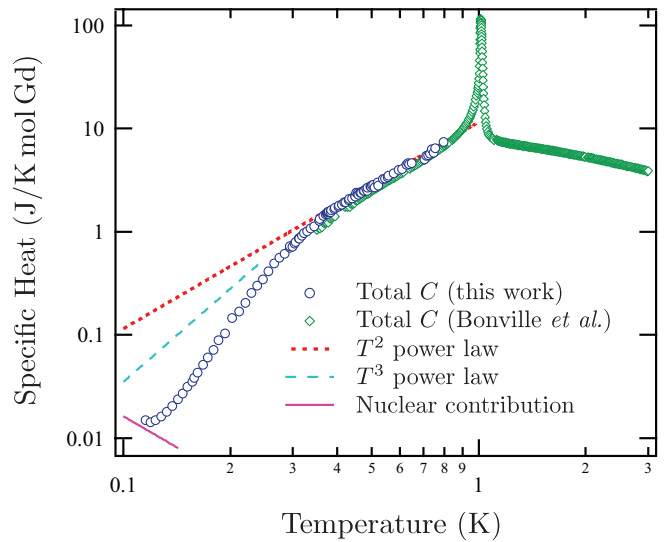}
\caption{\label{ComparisonPlot} 
(Color Online) Specific heat of Gd$_2$Sn$_2$O$_7$ as a function of temperature from this work
(blue circles) between 115 and 800 mK.  Previously measured data at higher temperatures
(green diamonds)~\cite{Bonville2003},  the $T^2$ power law previously proposed
  (dotted red line) and a $T^3$ power law (dashed blue line) are also plotted.
The upturn seen below 150 mK results from the nuclear electric quadrupole interaction (solid line).}
\end{figure}

Our specific heat results, shown in Fig.~\ref{ComparisonPlot}, are largely consistent with previous work~\cite{Bonville2003}, though there is a 10 to 15\% systematic discrepancy between the data sets.  Our measurements, however, extend to considerably lower temperature ($\sim 115$ mK) allowing us to test the proposal of gapped magnon excitations~\cite{Maestro2007}.  We observe below $\sim 350$ mK a deviation from the $T^2$ behavior describing the data between 350 mK and 800 mK as previously reported~\cite{Bonville2003}.  The specific heat decreases also faster than the $T^3$ power-law expected for conventional gapless antiferromagnetic magnons.  This already suggests that the specific heat may be dropping out exponentially, indicating a gapped spin-wave spectrum.

At the lowest temperatures an upturn in $C(T)$ becomes apparent and can be ascribed to a nuclear contribution, $C_N(T)$, stemming largely from the nuclear electric quadrupole interaction and, to a much lesser extent, to the nuclear hyperfine interaction.  By properly parametrizing this term and subtracting it from the total specific heat, we can isolate the contribution from the electronic moments.    $^{155}$Gd M\"{o}ssbauer spectroscopy experiments find a nuclear quadrupole electric splitting of -4.0 mm~s$^{-1}$~\cite{Barton1979,Bertin2002}.  More specifically, the four nuclear states are split with energies 0, 0.05, 12.1 and 15.9 mK as a result of both nuclear quadrupole and nuclear hyperfine interactions~\cite{Bertin2002}.  The only other naturally occurring isotope that contributes to the nuclear specific heat is $^{157}$Gd which, like $^{155}$Gd,  has spin $I=3/2$.  The ratio of the nuclear dipole moments is $\mu_N^{157}/\mu_N^{155} \simeq 0.8$ and the ratio of quadrupole moments is $Q_N^{157}/Q_N^{155} \simeq 1.1$~\cite{Speck1956}.  The quadrupole interaction is expected to be proportional to $Q_N I_z^2$ where the local [111] $z$-direction points to the center of the Gd tetrahedra.  There are no $I_x$ and $I_y$ terms in the nuclear electric quadrupole interaction due to axial symmetry of the Gd magnetic site~\cite{Barton1979}.  The nuclear hyperfine interaction is proportional to $\mu_N \vec{I}\cdot\vec{J}$. However, in the Palmer-Chalker ground state of GSO~\cite{Palmer2000}, the electronic moments are ordered perpendicular to the local $z$-direction, so we can simply write the nuclear hyperfine term proportional to $\mu_N I_x$.  Thus the total nuclear Hamiltonian is $\mathcal{H}_N = a\mu_N I_x + bQ_N I_z^2$.  We have calculated the nuclear contribution, $C_N(T)$, to $C(T)$ by choosing coefficients $a$ and $b$ such that the resulting eigenvalues of $\mathcal{H}_N$ match the above energy splittings for $^{155}$Gd.  The interactions are then scaled by the ratios of nuclear moments to get the contribution from $^{157}$Gd.  The calculated specific heats are multiplied by the relative natural isotopic abundances and added to give the total $C_N(T)$.

\begin{figure}
\includegraphics[width=3.375in,keepaspectratio=true]{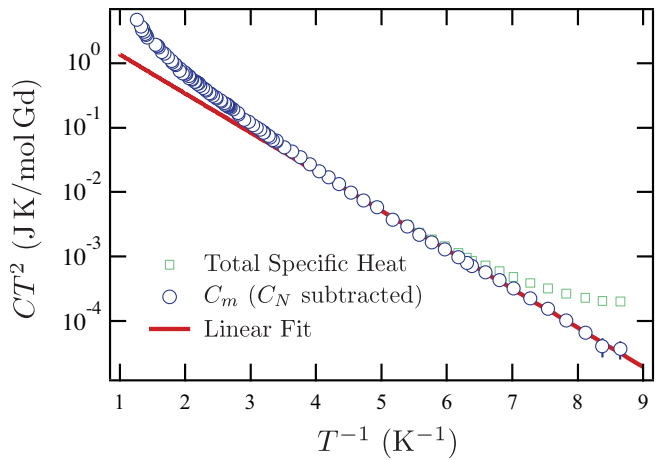}
\caption{\label{logCvs1overT} 
(Color online) $\log(CT^2)$ plotted versus $1/T$. The low temperature limit of the magnetic specific heat is parametrized by a phenomenological form $C_m\propto~(1/T^2)e^{-\Delta/T}$, which appears as a straight line.  The specific heat is plotted before (squares) and after (circles) subtraction of a nuclear contribution.  The linear fit is performed between the lowest temperature measured and 250 mK and gives $\Delta = 1.40\pm 0.01$ K.}
\end{figure}

At temperatures much higher than the nuclear energy splittings, the nuclear contribution behaves as $C_N = AT^{-2}$ with, based on the calculation above, $A\simeq 1.35\times10^{-4}$ JK/mol~Gd.  However, because of uncertainties in the nuclear moments, we keep $A$ as a free parameter.
$C_N(T)$ can be subtracted from the total $C(T)$ at these temperatures since the hyperfine coupling
is only a few mK.  As a first rough indicator of a
gapped excitation spectrum, we note that the electronic part of $C$ can be parametrized at low
temperature by $C_m(T) \propto (1/T^2)e^{-\Delta/T}$ as in conventional colinear antiferromagnets
with single ion anisotropy.  A plot of $\log(CT^2)$ versus $1/T$ should then give a straight line
and this appears to be the case if $A$ is chosen to be $1.63\times10^{-4}$ JK/mol~Gd as shown in
Fig.~\ref{logCvs1overT}, which is close to the value calculated above. The discrepancy is reasonable considering the uncertainty on the parameters in the calculation and the limited amount of data in the temperature range where the nuclear contribution becomes dominant. The phenomenological ``gap'' $\Delta$ so obtained is $\Delta=1.40 \pm 0.01$ K.  We now go on to describe a microscopic calculation which further supports the conclusion that a gapped magnon excitation spectrum exists in GSO.

Gd$_2$Sn$_2$O$_7$, and the related material Gd$_2$Ti$_2$O$_7$, are well approximated as isotropic, Heisenberg antiferromagnets as the Gd$^{3+}$ ions have half-filled shells ($L=0$, $S=7/2$).  Neutron scattering experiments have identified a $\vec{k}=(0,0,0)$ ordered ground state~\cite{Wills2006} -- the so-called Palmer-Chalker state~\cite{Palmer2000}.  Thus Gd$_2$Sn$_2$O$_7$ is particularly well suited to the standard protocol of, after having identified the  ground state, computing the second-quantized (magnon) low-lying spin excitations around that state~\cite{Maestro2004,Maestro2007}.  The nearest neighbor exchange interaction is estimated from the Curie-Weiss constant $\theta_{\rm CW} = -8.6$ K by $J_1 = 3\theta_{CW}/zS(S+1) = -0.273$ K~\cite{Bonville2003} where $z=6$ is the number of nearest neighbors. The strength of the dipolar interaction is derived from the size of the magnetic moments and the geometry of the lattice~\cite{Palmer2000,Wills2006,Maestro2004,Maestro2007}.  Though the moments are fairly isotropic, there exists a small crystal field anisotropy resulting from second order admixing from the electronic {\bf L}=0 state to excited manifolds~\cite{Glazkov2006}).  The crystal field parameter $B_2^0 = 47$ mK was used in this calculation as discussed in Ref.~\cite{Maestro2007}.

\begin{figure}
\includegraphics[width=3.375in,keepaspectratio=true]{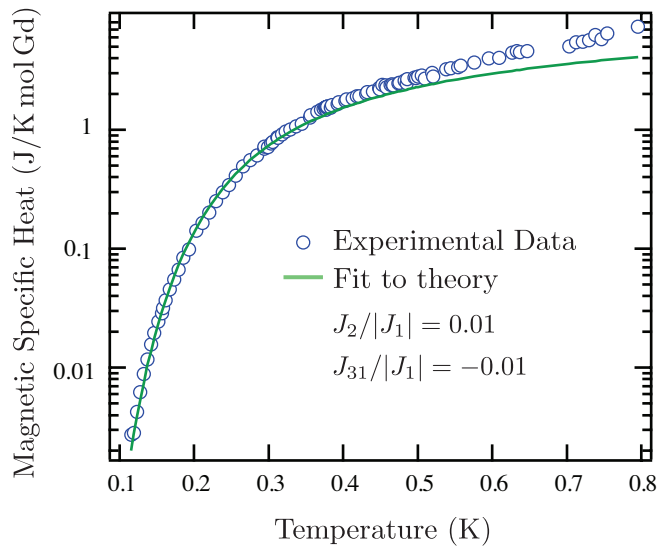}
\caption{\label{TheoryFit}
(Color online) Fit predicted by linear spin-wave theory to $C_m(T)$ 
with resulting values of the exchange interactions. The best fit gives $J_2/|J_1| = 0.01$ and $J_{31}/|J_1| = -0.01$.  Deviation from theory occurs  close to the transition, $T_c \simeq 1.0$~K, where magnon-magnon interactions become important.}
\end{figure}

A general two-body spin interaction Hamiltonian which includes isotropic magnetic exchange interactions up to third nearest neighbor and anisotropy in the form of interactions with the local crystal field as well as long range dipole-dipole interactions~\cite{Palmer2000,Wills2006,Maestro2004,Maestro2007}  was considered.  The techniques of linear spin-wave theory \cite{Maestro2004,Maestro2007} were employed to diagonalize the Hamiltonian and arrive at a Bose gas of non-interacting magnons created (annihilated) by operators $a^\dag_a(\vec{k})$ ($a^{\phantom\dag}_a(\vec{k})$),
\begin{equation}
\mathcal{H} = \mathcal{H}^{(0)} + \sum_{\vec{k}}\sum_a
\varepsilon_a (\vec{k})\left[
a^\dag_a(\vec{k})a^{\phantom\dag}_a(\vec{k}) + \frac{1}{2}\right],
\end{equation}
where $ \mathcal{H}^{(0)}$ is the classical ground state energy and $\varepsilon_a(\vec{k})$ are the spin-wave dispersion energies.  The pyrochlore lattice can be viewed as a FCC lattice with a 4-site basis (e.g. a Gd tetrahedron). The summation runs over all wavevectors $\vec{k}$ in the first Brillouin zone of the FCC lattice and the subscript $a$ runs over the four sublattice sites.  $C_m(T)$ is obtained from standard thermodynamic relations~\cite{Maestro2004,Maestro2007} 
\begin{equation}
C_m(T) = k_{\rm B} \beta^2
\sum_{\vec{k}}\sum_{a}[\varepsilon_a(\vec{k})]^2 
\frac{\exp[\beta\varepsilon_a(\vec{k})]}
     {[\exp[\beta\varepsilon_a(\vec{k})] - 1 ]^2}
\end{equation} 
where $\beta=1/k_{\rm B}T$, $k_{\rm B}$ the Boltzmann constant and $T$ the temperature.

Using a maximum likelihood estimator as in Ref.~\cite{Maestro2007}, we have confirmed that the specific heat is fit quantitatively well by a model that includes weak ferromagnetic second nearest neighbor and antiferromagnetic third nearest neighbor interactions ($J_2/|J_1| = 0.01$, $J_{31}/|J_1| = -0.01$) as seen in Fig.~\ref{TheoryFit}. Note, however, that qualitatively good fits are observed for a substantial region of the $J_2-J_{31}$ plane, and there can be little doubt about the presence of a gap to spin excitations at low temperatures.  With the above values of $J_2$ and $J_{31}$, the calculated dispersion relation shows a true spin wave gap, $\Delta_{\rm SW}\approx 1.24$ K, at the $\Gamma$ point  (${\bm k}=0$).  The difference between this $\Delta_{\rm SW}$ value and the $\Delta=1.40$ K value above arises from the fact that $\Delta$ is obtained on the basis of a phenomenological parametric form, $C_m(T)\propto(1/T^2)e^{-\Delta/T}$, while $\Delta_{\rm SW}$ is a truly microscopic gap.  With the values of $J_1$, $J_2$, $J_{31}$ above, the reduction of the classical order parameter $\langle S \rangle=7/2$ due to quantum fluctuations is $\langle \delta S\rangle \approx 0.11$~\cite{Maestro2004,Maestro2007}.


In conclusion, the experimental results presented here, in conjunction with those from a microscopic theoretical calculation that builds on a model that characterizes the long range ordered ground state of Gd$_2$Sn$_2$O$_7$~\cite{Palmer2000,Wills2006,Maestro2004,Maestro2007}, leave very little doubt that the bulk excitations in this system are conventional gapped collective magnons.  The following question therefore arises: what is the microscopic origin of the temperature-independent $\mu$SR relaxation rate~\cite{Bonville2004} and the higher than expected hyperfine temperature observed with M\"{o}ssbauer spectroscopy~\cite{Bertin2002}?  Such gapped spin-wave excitations would typically be expected to lead to a sharply dropping spin relaxation rate.

The peculiar $T^2$ behavior of the specific heat of GTO also remains a mystery for, in contrast to GSO, it does not exhibit a gap down to 100 mK~\cite{Yaouanc2005} as might be naively expected.  It would be valuable to investigate the possibility that the $T^2$ power law in GTO also gives way to an exponentially decaying specific heat at lower temperature. GSO and GTO are perhaps at this time the two pyrochlore materials most amenable to matching theory with experiment~\cite{Maestro2004,Maestro2007}. Hence determining the cause of PSDs in these pivotal systems may provide insight into the cause of persistent spin dynamics in other more exotic, highly frustrated systems such as Tb$_2$Ti$_2$O$_7$~\cite{Gardner1999}, Tb$_2$Sn$_2$O$_7$~\cite{Bert2006}, Yb$_2$Ti$_2$O$_7$~\cite{Hodges2002} and Gd$_3$Ga$_5$O$_{12}$ (GGG)~\cite{Dunsiger2000}.


\begin{acknowledgments}

We thank Rob Kiefl for stimulating discussions. Funding for this research was provided by NSERC, CRC (Tier 1, M.G.), CFI, MMO and Research Corporation.  We thank P. Bonville for kindly providing us with specific heat data for the purpose of comparison. 
M.G. acknowledges the U. of Canterbury (UC) for support and the hospitality of the Dept. of Physics and Astronomy at UC where part of this work was completed.

\end{acknowledgments}



\end{document}